\documentclass[12pt,preprint]{aastex}

\begin{document}
\title{The optical flare and afterglow light curve of GRB 050904 at redshift z=6.29}

\author{D. M. Wei$^{1,2}$, T. Yan$^{1,2}$, and
 Y. Z. Fan$^{1,2}$}
\affil{$^1$ Purple Mountain Observatory, Chinese Academy of Science,
Nanjing 210008, China.\\
$^2$ National Astronomical Observatories,Chinese Academy of
Sciences, Beijing, 100012, China.}

\begin{abstract}
GRB050904 is very interesting since it is by far the most distant
GRB event known to date ($z=6.29$). It was reported that during
the prompt high energy emission phase, a very bright optical flare
was detected, and it was temporal coincident with an X-ray flare.
Here we use two models to explain the optical flare, One is the
``late internal shock model", in which the optical flare is
produced by the synchrotron radiation of the electrons accelerated
by the late internal shock, and the X-ray flare is produced by the
synchrotron-self-Compton mechanism. The other is the external
forward-reverse shock model, in which the optical flare is from
the reverse shock emission and the X-ray flare is attributed to
the central engine activity. We show that with proper parameters,
a bright optical flare can appear in both models. We think the
"late internal shock model" is more favored since in this model
the optical flash and the X-ray flare have the same origin, which
provides a natural explanation of the temporal coincidence of
them. In the forward-reverse shock scenario, fits to the optical
flare and the late afterglow suggests that the physical parameters
of the reverse shock are much different from that of forward
shock, as found in modeling the optical flash of GRB 990123
previously.

\end{abstract}


\section{Introduction}
Gamma-Ray Bursts (GRBs) are bright flashes of high energy photons
usually lasting about several seconds. They are by far the most
luminous objects in the universe, they emit so large amounts of
energy (up to $10^{53}$ ergs) and thus can be detected up to very
high redshifts ($z>5$).

GRB050904 was detected by the Burst Alert Telescope (BAT) onboard
Swift on 2005 September 4 at 01:51:44 UT (Cummings et al. 2005). It
was a long ($\leq 500$ seconds duration in BAT), multi-peaked, bight
burst, the 15 - 150 keV fluence was ($5.4 \pm 0.2$) $\times 10^{-6}$
erg cm$^{-2}$, the spectrum can be described by a power law with a
photon index $\sim -1.34$, its redshift has been measured by several
groups (Haislip et al. 2005; Antonelli et al. 2005; Price et al.
2005), $z=6.29$ makes it to be by far the most distant GRB
discovered to date.

Bo\"{e}r et al. (2005) reported that they detected a very bright
optical flare during the prompt high energy emission phase, and at
the same time there is an X-ray flare. It is widely believed that
the reverse shock synchrotron radiation usually peaks in the
optical/IR band, and this emission has been successful in
interpreting the early optical emission from GRB990123 (Akerlof et
al. 1999; Sari \& Piran 1999; Wang et al. 2000; Fan et al. 2002;
Zhang et al. 2003; Nakar \& Piran 2005), GRB021211 (Fox et al.
2003; Li et al. 2003; Wei 2003; Kumar \& Panaitescu 2003),
GRB041219a (Blake et al. 2005; Fan et al. 2005b) and GRB 050525a
(Blustin et al. 2005; Shao \& Dai 2005). However in this reverse
shock model, it is expected that the emission has a negligible
contribution in the X-ray band (However, see Fan \& Wei 2005).
Strong optical flare accompanying an X-ray flare may also be
accounted for by the ``late internal shock model" (Fan \& Wei
2005). Originally, that model has been proposed to interpret the
X-ray flare detected in GRB 011121 (Piro et al. 2005) and many XRT
X-ray flares (Burrows et al. 2005; Zhang et al. 2005; Nousek et
al. 2005).

The optical afterglow light curve of GRB050904 cannot be described
by a simple power law, between $\sim 3$ hours and 0.5 day after
the burst, the fading of the afterglow can be described by a power
law with index -1.36. However after this time the light curve
flattened to a temporal index of -0.82 (Haislip et al. 2005).
Tagliaferri et al. (2005) have found a break in the light curve at
time $t_b\simeq2.6$ day, which may be the jet effect. In this
Letter, we try to explain the optical flare with two models, i.e.
the reverse shock emission and the late internal shock model, and
then we will fit the afterglow light curve including energy
injection and jet effects.

\section{Explanation of the optical flare}

\subsection{The late internal shock model}
In the standard shock scenario, the prompt gamma-ray emission is
produced by the internal shock, and the burst duration is determined
by the active timescale of the central engine. However, some authors
suggest that the activity of the central engine may be much longer
than the GRB duration, which can give rise to some signatures in
multi-wavelength afterglows (Dai \& Lu 1998; Zhang \& M\'esz\'aros
2001; Granot et al. 2003; Ioka et al. 2005). Furthermore, it has
been proposed that the Fe line observed in some GRB X-ray afterglows
are produced by the late time energy injection (Rees \& M\'esz\'aros
2000; Gao \& Wei 2005).

Fan \& Wei (2005) first proposed the late internal shock model to
account for the bright X-ray flares detected in many GRBs. Here we
will show that the late internal shock model not only can produce
the X-ray flare, but also can produce the optical flare with
proper parameters.

Following Fan \& Wei (2005), the typical synchrotron radiation
frequency can be estimated by

\begin{eqnarray}
\nu_{m} & \approx & 8.5\times 10^{15} (\frac{\epsilon_e}{0.4})^2
\epsilon_{B,-2}^{1/2} (\Gamma_{sh}-1)^{5/2}\Gamma_{sh}^{1/2}
L_{m,52}^{1/2}\nonumber\\
&& \Gamma_2^{-2}\delta t_{1}^{-1} ~{\rm Hz},
\end{eqnarray}

where $L_m$ is the outflow luminosity, $\Gamma_{sh}$ is the Lorentz
factor of the internal shock, $\Gamma$ is the Lorentz factor of the
emitting shell, $\delta t$ is the observed typical variability
timescale, $\epsilon_B$ and $\epsilon_e$ are the energy fractions
occupied by the magnetic field and electrons, respectively. Here the
convention $Q_x=Q/10^x$ has been adopted in cgs units throughout the
text.

The cooling Lorentz factor is $\gamma_{e,c}\simeq 7.7\times 10^{8}
(1+z)/[(1+Y)\Gamma B^2 \delta t]$, where
$Y=[-1+\sqrt{1+4x\epsilon_e/\epsilon_B}]/2$ is the Compton
parameter, $x\simeq \min\{1, (\nu_m/\nu_c)^{(p-2)/2}\}$ (Sari \&
Esin 2001). Then the cooling frequency is

\begin{eqnarray}
\nu_{c} &\approx & 1.6\times 10^{11}(\frac{1+z}{7.29})^{-2}
\epsilon_B^{-3/2}[\Gamma_{sh}(\Gamma_{sh}-1)]^{-3/2}
L_{m,52}^{-3/2}\nonumber\\
&& \Gamma_2^8 \delta t_{1} (1+Y)^{-2} ~{\rm Hz}
\end{eqnarray}

The synchrotron-self-absorption frequency is about (Li \& Song
2004; Fan \& Wei 2005)

\begin{eqnarray}
\nu_{a} &\approx & 2.9\times 10^{14} (\frac{1+z}{7.29})^{-2/7}
\epsilon_{B,-2}^{1/14} [\Gamma_{sh}(\Gamma_{sh}-1)]^{1/14}
L_{m,52}^{1/14}\nonumber\\
&& L_{syn,50}^{2/7} \Gamma_2^{-8/7}\delta t_{1}^{-5/7} ~{\rm Hz}
\end{eqnarray}

where $L_{syn}$ is the synchrotron radiation luminosity. The
maximum flux of synchrotron radiation is $F_{max}\approx
3\sqrt{3}\Phi_p (1+z)N_e m_e c^2 \sigma_T \Gamma B/(32\pi^2 q_e
D_L^2)$, where $q_e$ is the charge of electron, $N_e=L_m \delta
t/[(1+z)\Gamma m_p c^2$ is the total number of emitting electrons,
$\Phi_P$ is a function of $p$, for $p=2.5$, $\Phi_P \simeq 0.6$
(Wijers \& Galama 1999). $D_L$ is the luminosity distance, we
adopt $(\Omega_M,\Omega_\Lambda,h)=(0.3,0.7,0.71)$. Then for the
case $\nu_{c}< \nu_{a}< \nu_{obs} < \nu_{m}$, the observed flux at
frequency $\nu_{obs}$ should be

\begin{eqnarray}
F_{\nu} &\approx & 100(\frac{\nu_{obs}}{3\times 10^{14}{\rm
Hz}})^{-1/2} L_{m,52}^{3/4}\Gamma_2 \epsilon_{B,-2}^{-1/4}
[\Gamma_{sh}(\Gamma_{sh}-1)]^{-1/4}\nonumber\\
&& D_{L, 29.3}^{-2} \delta t_{1}^{1/2}(1+Y)^{-1} ~{\rm mJy}
\end{eqnarray}

Now we turn to the observation. Bo\"{e}r et al. (2005) reported
that they detected a bright optical flare at frequency $\nu_{obs}
=3\times 10^{14}$ Hz, the peak flux is 48 mJy. Meanwhile, the
Swift XRT data shows that there is also a peak in the X-ray light
curve at nearly the same time with the optical flare, which
suggests that the optical flare and the X-ray peak may have the
same origin. The slope of the X-ray spectrum is about -1/2, and
the flux at $1 {\rm KeV}$ is about 0.08 mJy.

In our late internal shock model, if we take the values as
follows: $\epsilon_e=0.4$, $\epsilon_B=0.02$, $L_m=10^{52}$
ergs$^{-1}$, $\Gamma=200$, $\Gamma_{sh}=1.6$, $\delta t =20s$,
then we find $\nu_{m}\sim 6.3\times 10^{14}$ Hz, $\nu_{a}\sim
8.4\times 10^{13}$ Hz, $\nu_{c}\sim 1.2\times 10^{12}$Hz, so it is
in the fast cooling phase, between $\nu_{a}$ and $\nu_{m}$ the
spectrum takes the form $F_{\nu}\propto \nu^{-1/2}$, and at the
observed frequency ($3\times 10^{14}$ Hz) the flux is 49 mJy,
which is quite consistent with the observation. In addition, with
the values of $\epsilon_e$ and $\epsilon_B$, the Compton parameter
$Y\simeq 4$, then the synchrotron photons will be Compton
scattered to high energy, the energy spectrum between $10^{16}$ Hz
and $10^{19}$ Hz is also $F_{\nu}\propto \nu^{-1/2}$, and we can
estimate the flux at $1 {\rm KeV}$ to be about 0.06 mJy, which is
also consistent with the observation well.

\subsection{The reverse shock model}
\label{sec:OFRS} After the internal shock phase, as the fireball
is decelerated by the circumburst medium, usually a pair of shocks
develop (M\'esz\'aros \& Rees 1997; Sari \& Piran 1999; Kobayashi
2000). The early optical afterglow lightcurve is usually composed
of the contributions from both the forward (FS) and the reverse
shocks (RS). With that model, the very early optical/IR flash
following GRB 990123, GRB 021211, GRB 041219a and GRB 050525a
could be well modelled by assuming the physical parameters are
quite different for the FS and RS (Fan et al. 2002; Zhang et al.
2003; Kumar \& Panaitescu 2003; McMahon et al. 2004; Fan et al.
2005b; Blustin et al. 2005). For example, Fan et al. (2002)
performed a detailed fit to the optical flash of GRB 990123 data
and obtained $\epsilon_{e} ^{\rm r}=4.7\epsilon_{e}^{\rm f}=0.6$
and $\epsilon_{ B}^{\rm r}=400\epsilon_{B} ^{\rm f}=0.4$, where
the superscripts ``r'' and ``f'' represent RS and FS,
respectively.

B\"oer et al. (2005) found that both the optical flare and the
gamma-ray burst of GRB 050904 were as energetic as those of GRB
990123 (in the rest frame of the GRBs). If other parameters
(including the initial Lorentz factor of the ejecta, the number
density of the interstellar medium $n$) are similar for these two
events, then the resulted shock parameters should be similar, too!
So it is very likely that in the current case the shock parameters
of FS and RS are also different.

Recently, Yan et al. (2005) have developed a code to calculate the
GRB afterglow light curves, including the FS and the RS emission
components. In the current calculation, there are two novel effects
have been taken into account. One is that in previous works, the
Lorentz factor of the outflow as well as the comoving number density
of particle are assumed to be constant. This may not be the case
since in the standard fireball model, the gamma-ray burst is from
the internal shocks. The detected gamma-ray lightcurve is so
variable that the involved outflow may be variable, too (both the
Lorentz factor and the particle number density). In order to model
the optical plateau (Bo\"er et al. 2005) and partly for convenience,
in this work we assume the outflow can be approximated as two parts.
Their bulk Lorentz factors, isotropic energies and widths are
($\eta_{(1)}$, $E_{iso(1)}$, $\Delta_{(1)}$) and ($\eta_{(2)}$,
$E_{iso(2)}$, $\Delta_{(2)}$), respectively. The other is the more
reliable calculation of the arriving time of the RS emission. We
take the emitting time of the first $\gamma-$ray photon as our zero
point of time. On the line of sight ($\theta=0$), a gamma-ray photon
$\gamma_P$ arriving us at $t$ implies that the distance of
corresponding electron (i.e., point P, at which the bulk Lorentz
factor is $\eta$) to the initial outflow front is $\sim c t/(1+z)$.
The radial distance of the FS front to the central engine is $R_P$
when the RS crosses point P. At that time, the width between photon
$\gamma_P$ and point P is $\approx (1-\beta_\eta)R_P$, where
$\beta_\eta=\sqrt{1-1/\eta^2}$. Therefore, the arriving time of the
RS emission from point P should be $t+(1+z)(1-\beta_\eta)R_p/c$. It
is straightforward to extend this calculation to the cases of
$\theta \neq 0$. It is found that the $I-$band flare of GRB 050904
can be well reproduced with the following parameters (see the insert
of Fig. \ref{fig:Fits}): the isotropic energy of the outflow is
$\eta_{(1),2}=380$, $E_{iso(1),54}=0.4$, $\Delta_{(1),12}=1.3$,
$\eta_{(2),2}=800$, $E_{iso(2),54}=0.3$, $\Delta_{(2),12}=0.7$,
$n=3~{\rm cm^{-3}}$, $\epsilon_e^{\rm r}=0.6$, and $\epsilon_B^{\rm
r}=0.4$. It is surprising to see that the resulting reverse shock
parameters are nearly the same as those of GRB 990123 (Fan et al.
2002).

Can the X-ray flare be from the RS, too? The answer is negative.
Firstly, as shown by Fan \& Wei (2005), the decline of the X-ray
emission of the RS can not be steeper than $t^{-(2+p/2)}$, which
is inconsistent with the observation. Secondly, now the reverse
shock region is significantly magnetized, so the RS emission in
X-ray band should also be dominated by the synchrotron radiation.
Thus the X-ray band emission should be an extension of the optical
emission. However, the observation shows the optical-to-X-ray
bands emission can not be described by a simple synchrotron
spectra (Bo\"er et al. 2005). Therefore, the X-ray flare
accompanying the optical flare should be attributed to the
activity of the central engine in the RS model.

\section{Fits to the late $J-$band afterglow}
\label{sec:LateAfterglow} The multi-wavelength afterglow light
curves (especially the $J-$band one) of GRB050904 have been
detected (Haislip et al. 2005; Tagliaferri et al. 2005  and the
references therein). Between $\sim 3$ hours and 0.5 day after the
burst, the fading of the afterglow can be described by a power law
with index -1.36. After that time the light curve flattened to a
temporal index of -0.82. A break appears at time $t_b\simeq2.6$
day, which suggests that the outflow may be a jet. In this Letter
we pay more attention to the optical flattening. We note that at
the observer time $t\sim 10^4-10^5$ s, there are strong X-ray
flares (Cusumano et al. 2005; Price et al. 2005; Watson et al.
2005). Fan \& Wei (2005) suggested that when the moderate
relativistic outflow powering the X-ray flare catched up with the
initial GRB ejecta, a flattening would occur in the
long-wavelength afterglow light curve. In the calculation, we
assume that between $t\sim 4\times 10^4$ seconds and $t\sim 1.5$
day, significant part of energy has been injected into the
decelerating GRB ejecta. Similar to Zhang et al. (2005), the
energy injection rate has been taken as $dE_{inj}/[dt/(1+z)]=
Ac^2(t/t_0)^{-0.5}$, where $A$ is a constant. We take $A=0$ when
there is no energy injection. With the energy injection, the
equation (8) of Huang et al. (2000) should be replaced by

\begin{equation}
d \gamma={(1-\gamma^2)dm+{A(t/t_0)^{-0.5}[dt/(1+z)]}\over
M_{ej}+\epsilon m+2(1-\epsilon)\gamma m},
\label{Eq:Dyn}
\end{equation}
where $\gamma$ is the bulk Lorentz factor of the GRB ejecta,
$M_{ej}$ is the rest mass of the initial GRB ejecta, $m$ is the
mass of the medium swept by the GRB ejecta (which is governed by
$dm=4\pi R^2 n m_p dR$, where $m_p$ is the rest mass of proton,
$dR=\gamma(\gamma+\sqrt{\gamma^2-1})c dt/(1+z)$, $\epsilon=x
\epsilon_e$ is the radiation efficiency. With the dynamical
evolution of the ejecta, it is straightforward to calculate its FS
emission (e.g., Huang et al. 2000; Yan et al. 2005).

\begin{figure}
\epsscale{1.0} \plotone{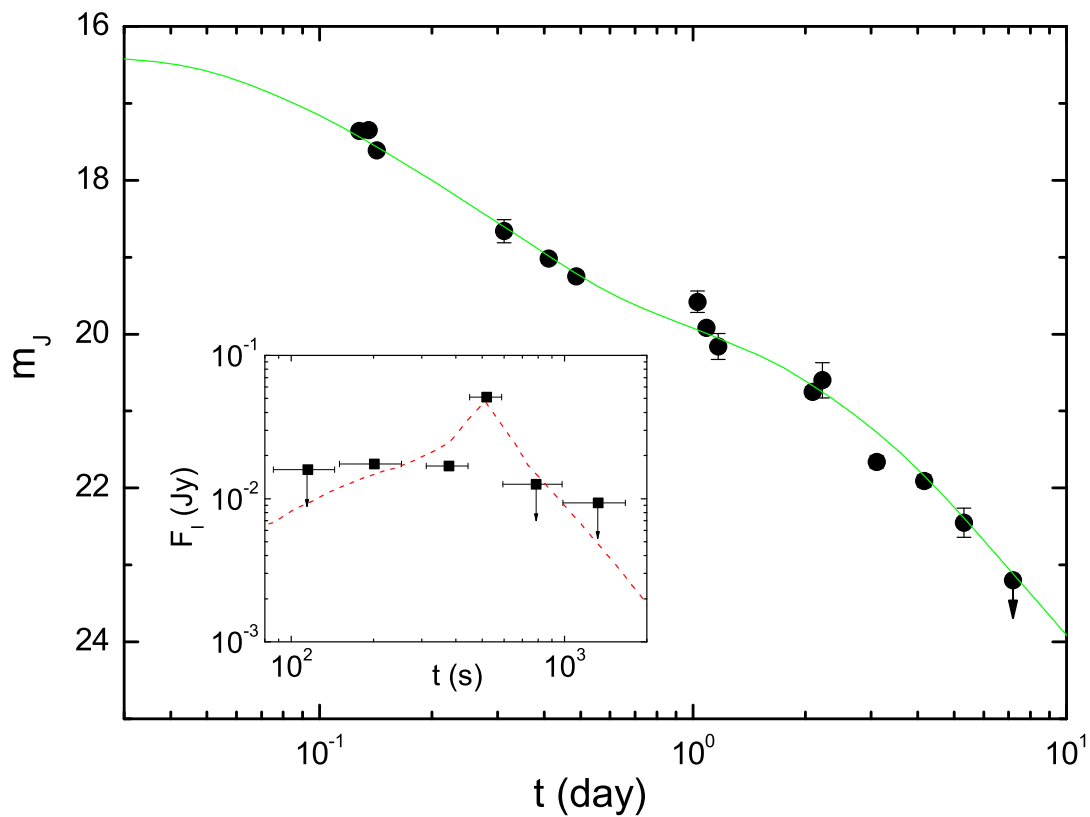} \caption{Modeling the $I$-band
flare (the insert) and the $J$-band afterglow of GRB 050904. In
the insert, the $I-$band flare data (the filled rectangles) are
taken from Bo\"er et al. (2005), the dashed line is the
theoretical light curve of the reverse shock emission. The
$J-$band afterglow data (the filled circles) are taken from
Haislip et al. (2005) and Tagliaferri et al. (2005). The solid
line is the theoretical light curve of the $J-$band afterglow.}
\label{fig:Fits}
\end{figure}

The fits to the $J-$band data (taken from Haislip et al. [2005] and
Tagliaferri et al. [2005]) are presented in Fig. \ref{fig:Fits}. It
is found that the data can be well modeled with the following
parameters: $E_{\rm iso,54}=0.7$, $n=3~{\rm cm^{-3}}$,
$\epsilon_e^{\rm f}=0.15$, $\epsilon_B^{\rm f}=0.001$, $A=7\times
10^{49}~{\rm ergs~s^{-1}}$, $t_0=4\times 10^4$ s, and the jet angle
$\theta_j=0.054$. Note that the value of $\theta_j$ is obtained from
fitting the afterglow light curve, not from the simple analytic
relation. Comparing with the reverse shock parameters derived in
\S{\ref{sec:OFRS}}, the shock parameters of the FS and the RS are
quite different, as that found in GRB 990123 (Fan et al. 2002; see
also Zhang et al. 2003). The isotropic energy of the $\gamma-$rays
is $\sim 5\times 10^{53}$ ergs and the derived $\theta_j=0.054$, so
the geometry corrected energy should be $\sim 7\times 10^{50}$ ergs,
which is typical for the GRBs detected by BeppoSAX, HETE-2 and
Swift. In our treatment, the flattening is caused by the late time
energy injection. The total isotropic energy injected into the GRB
ejecta is $\sim 6\times 10^{53}$ ergs.

\section{Discussion and conclusion}
The bright optical flare has been detected in GRB050904, which is as
bright as the optical flash of GRB990123 (in the rest frame of
bursts) and seems to be accompanied by an X-ray flare (Bo\"er et al.
2005). Here we explored two possible models to account for that
observation. One is the ``late internal shock model", in which the
optical flare is produced by the synchrotron radiation of the
electrons accelerated by the late internal shock, and the X-ray
flare is produced by the synchrotron-self-Compton process.
\footnote{However, in some cases the synchrotron emission may peak
in keV energy band, then the IC component would peak at GeV energy
band (unless the outflow is highly magnetized, as suggested by Fan,
Zhang \& Proga (2005a)), which may be detectable for the upcoming
GLAST.  This possibility will be discussed in great detail
elsewhere.}
 The other is the external forward-reverse shock model, in which the
optical flare is from the reverse shock emission and the X-ray
flare is attributed to the central engine activity.  We show that
with proper parameters, a bright optical flare can appear in both
models.

In the forward-reverse shock scenario and with late time energy
injection, we have modeled the optical flare as well as the late
$J-$band afterglow numerically. The resulted shock parameters of
the forward/reverse shocks are $\epsilon_e^{\rm
r}=4\epsilon_e^{\rm f}=0.6$ and $\epsilon_B^{\rm
r}=400\epsilon_B^{\rm f}=0.4$, respectively. They are quite
similar to those found in GRB 990123 (Panaitescu \& Kumar 2001;
Fan et al. 2002), which is a natural result in view of the
similarity between these two GRBs and their optical flares (in the
rest frame of bursts).

As for the reverse shock emission, previous works usually assumed
that the physical parameters are uniform, this greatly simplify the
calculation. But in reality, the observed gamma-ray emission light
curve is much variable, so it is very likely that the involved
outflow should be also variable. We notice that if the parameters
are uniform, then before the peak time the flux rises quickly,
cannot account for the observed plateau (Kobayashi 2000; Bo\"{e}r et
al. 2005). Here just for simplicity, we divide the outflow into two
parts. We expect that in the realistic case, the outflow should be
non-uniform, so the parameters should have a continuous distribution
within the shell, but the calculation is complicated.

Although both the "late internal shock model" and the reverse shock
emission can account for the observed optical flash and the X-ray
flare, We think the "late internal shock model" is more favored
since in this model the optical flash and the X-ray flare have the
same origin, which provides a natural explanation of the temporal
coincidence of them. In the late internal shock model, it is needed
that after the prompt $\gamma-$ray burst phase, the central engine
could re-start. Recently two models have been proposed for the
production of late energy injection (King et al. 2005; Perna et al.
2005; MacFadyen et al. 2005 proposed another model to account for
the X-ray flare in short GRBs). While in the reverse shock model,
the temporal coincidence of the optical flash and the X-ray flare
can only be regarded as fortuitous. In addition, we note that in the
late internal shock model, the typical synchrotron radiation
frequency strongly depends on the parameters, such as $\Gamma$,
$\Gamma_{sh}$, $L_m$, $\delta t$ etc., and for different burst
sources it is natural that these parameters are different,so we
expect that the late internal shock model not only can produce the
optical or X-ray flare, but also can produce the flare at other
wavelength, such as at the ultraviolet or infrared. Meanwhile we
predict that the synchrotron-self-Compton process may produce
emission at high energy band ($\sim$ GeV).

Despite its high redshift, the optical afterglow of GRB050904 is not
peculiar with respect to other GRBs. Recently Zhang et al. (2005)
and Nousek et al. (2005) analyzed the X-ray afterglows of many GRBs,
they found some features (the X-ray flares, the flattening of the
light curve, a late time break) occurred in a good fraction of GRBs.
These features are consistent with the afterglow of GRB050904. We
suggest that the progenitor of GRB050904 may be not quite different
from that of other GRBs in view of these similarity.

\acknowledgments We thank the referee for her/his helpful comments.
This work is supported by the National Natural Science Foundation
(grants 10225314 and 10233010) of China, and the National 973
Project on Fundamental Researches of China (NKBRSF G19990754).

\end{document}